\documentclass[12pt,preprint]{aastex}

\shorttitle{GRB Morphologies}
\shortauthors{Hakkila and Giblin}

\begin{document}

\title{A Simple Two-Parameter Characterization \\
    of Gamma-Ray Burst Time Histories}


\author{Jon Hakkila and Timothy W. Giblin}
\affil{Department of Physics and Astronomy, College of Charleston, 
Charleston, SC  29424-0001}







\begin{abstract}
A simple scheme delineates Long GRBs with similar time histories using the Internal Luminosity Function power-law index and the spectral lag. Several generalizations are made about time history morphologies: 1) GRBs with long spectral lags contain fewer pulses that are broader than those found in bursts with short spectral lags, 2) short-lag bursts with small ILF power-law indices have many narrow pulses and are highly variable, while long-lag bursts with small ILF power-law indices are characterized by broad, smooth pulses and have low variability. 

GRB time history morphologies primarily identify intrinsic rather than extrinsic characteristics based on correlations with gamma-ray luminosity, afterglow luminosity, and numbers of pulses. These characteristics result because internal relativistic effects due to bulk Lorentz factor are larger than cosmological effects, and because the numbers and shapes of pulses indicate different efficiencies and forms of GRB energy release. 

Single-pulsed GRBs are characterized either by large ILF power indices (indicating a range of jet opening angles and Lorentz factors, with a FRED pulse shape), or they have long lags (large jet opening angles with low Lorentz factors, with either a FRED pulse shape or an unpeaked, smooth pulse shape). They also have lower-luminosity afterglows than multi-pulsed GRBs. GRBs with simple time histories are often associated with Type Ibc supernovae. This suggests that some single-pulsed GRBs contain single, beamed blast waves that are similar to and have characteristics that overlap with those of many supernovae. Such a connection may not exist between multi-pulsed GRBs and supernovae.

\end{abstract}

\section{Introduction}

Gamma-ray burst (GRB) time histories are often unique and 
difficult to characterize; this makes them difficult to predict from other 
measured burst properties. GRB temporal structures and spectral characteristics are complex 
and evolve significantly over short timescales, resulting in a large variation from burst to burst.  Still, some consistent burst morphologies have been identified. The most easily recognized of these is the Short class of GRBs. Although these bursts are identified on the basis of duration and spectral hardness (e.g. \citet{kou93, muk98,hak03}), they are found to have short, hard time histories with millisecond fluctuations.
Also recognized are `FRED' (Fast Rise Exponential Decay) bursts \citep{mee91}; these are characterized by a rapid rise with a long, smooth decay. 
Investigators also often allude to `complex' or `spiky' bursts; these multi-peaked bursts 
can vary rapidly and often have a hard spectral component.
We would like to find ways of sorting GRB time histories into meaningful morphologies so that we can better understand their physical properties.

GRB time histories identify kinetic energy dissipation from a relativistic particle flow. It is generally accepted that relativistic shocks dissipate this flow, but other dissipation methods are possible (e.g., see the review article by \citet{pir05}). Since the flow is beamed emission, these records are not complete observations, but only contain the portions of the dissipation that is beamed in the direction of the observer. The set of observed GRB time histories thus represents a sketchy and incomplete record; a skeleton from which we must construct a fleshed-out understanding of GRB physics.

Pulses seem to be the basic component of GRB prompt emission \citep{for95, lia96, nor96, bor01}.  Although this is a very useful key concept and building block, the delineation and study of GRB pulses is difficult and has primarily been performed only for bursts with non-overlapping pulses \citep{kar95, nor96, pen97, lee00, rrm00, sca00, koc01, ryd02}. We have identified a simple two-parameter scheme for sorting GRB time histories into physically meaningful patterns that relate to GRB pulses. 

\section{Data}

The two attributes used 
in this simple scheme are the {\em energy-dependent spectral lag} and the power-law index of {\em the 
Internal Luminosity Function}. The database used in obtaining these attributes is the 64 ms 
GRB database collected by the Burst And Transient Source Experiment (BATSE) on 
NASA's Compton Gamma Ray Observatory.

\subsection{Spectral Lags}

Spectral lags refer to the time delay between hard and soft gamma-ray prompt emission in GRBs. 
Lags are obtained from the peak of the cross correlation function or CCF \citep{ban97}
between two energy channels. From BATSE's four broadband energy
channels (Channel 1 between 20 and 50 keV, Channel 2 between 50 and 100 keV, Channel 3
between 100 and 300 keV, and Channel 4 between 300 keV and roughly 1 Mev), six different lag measurements can be obtained for each GRB. 
Since the peak of the cross-correlation function can be difficult to
identify due to background fluctuations, a nonlinear fit is made to the 
distribution in order to determine the statistical location of the peak. We have chosen to use 
a GRB pulse model \citep{nor96} fit to the CCF rather than a quadratic \citep{wu00} or 
a cubic \citep{nor00}. Although this time-asymmetric function is not as simple as a cubic, its additional degrees of freedom can result in a more accurate CCF fit. Also, the model diverges less from the measured data away from the pulse peak where the pulse shape changes dramatically. \citet{wu00} and \citet{nor00} have demonstrated that the fitted CCF range affects the lag measurement; an improvement can be made to the fit using data far from the CCF peak unless the pulse shape deviates from the fitting function at these values. This dilemma can be reduced by averaging lags obtained from CCF measurements spanning a range of temporal shifts (typically, 5 to 8 trial measurements made over a broad range of CCF values in the vicinity of the CCF peak). We note that these trials are not independent, so the lag error can be approximated by the average error obtained from each individual trial.  All lags shorter than 64 ms are difficult to measure accurately, since 64 ms is the minimum bin size available from this dataset.

We have chosen to use the channel 31 lag measurement ($\rm lag_{31}$) because 
1) it spans a large energy range and thus produces a long, easy-to-measure lag, 
2) channel 4 emission is not detected for a large number of BATSE GRBs, and 
3) bursts usually peak in channel 3, which increases the signal-to-noise ratio needed to 
measure the CCF. 

The distribution of $\rm lag_{31}$ measurements is found to have a peak near zero 
seconds. It appears likely from our analysis that few GRBs are actually characterized 
by negative lags; the lag distribution appears to be generally consistent with positive values 
and measurement errors that are large compared to the 64 ms temporal window size. 
This is consistent with the result of \citet{nor02}, who showed
that Long GRBs have positive lags and that negative lag measurements are due to
bursts with lower signal to noise (Short GRBs
have lags near zero as measured by both BATSE and Swift \citep{nor06}).
We therefore choose to exclude GRBs with zero or negative lags from our analysis, attributing them to
measurement error, a separate burst class, and the inability to plot bursts with 
$\log(\rm lag_{31}) = -\infty$. The uncertainty in our lag measurements $\sigma_{\rm lag_{31}}$ 
can be roughly approximated by the relationship
$\sigma_{\rm lag_{31}} \approx 0.12 \sqrt{\rm lag_{31}}$, where both $\rm lag_{31}$ and $\sigma_{\rm lag_{31}}$ are measured in units of seconds. This relationship is an empirical fit to the lags of 1482 BATSE GRBs: it demonstrates that the lag uncertainty scales with the lag (and therefore with the pulse width), but does not attempt to correlate the lag with peak intensity other than the observed anti-correlation between pulse width and burst peak intensity.

\subsection{Internal Luminosity Function}

The internal luminosity function $\psi(L)$ (or ILF) is the distribution of luminosity within 
a gamma-ray burst: $\psi(L)dL$ represents the fraction of time during which a burst's 
luminosity lies between $L$ and $L+dL$ \citep{hor97}.
We calculate the ILF using concatenated BATSE 64-ms data from HEASARC's Compton 
Gamma-Ray Observatory Science Support Center. A detailed description of the process used 
will be described in greater detail \citep{hak06}, but a summary of it is given
here. The ILF is measured for numerous combinations of BATSE's four broadband energy 
channels as follows: 1) A representative background level is obtained during the burst 
for the purpose of estimating average Poisson fluctuations. 2) A fit is obtained to model
time-dependent background variations, which are then removed. 3) Occasionally, some burst 
time intervals have poor time resolution (e.g. a small number of GRBs have pulses in the 
1024 ms pre-trigger data). Monte Carlo models of Poisson variations are used to artificially create 
similar temporal resolution in these time intervals.  
4) A distribution function is constructed by binning count rates relative to a defined 
minimum (e.g. $1\sigma$, $2\sigma$, or $3\sigma$ above the background). 
Expected Poisson background rates are subtracted from each ILF bin so that only estimated
source counts remain. This approach allows the technique to be applied to flux contributions
that are fainter than the $3\sigma$ limit used by \citet{hor97}, and therefore allows the
technique to be applied to many dim bursts. 
5) Initially, a large number of bins is chosen, similar to the number of bins used by 
\citet{hor97}. For fainter and/or shorter bursts, the number of bins is reduced in an iterative fashion until each bin contains at least five measurements (a minimum number assumed for Gaussian statistics). 
6) The ILF is normalized by the requirement that 
$\Sigma\psi(L)\Delta L=1$.

The ILF can generally be fitted by a quasi power-law form such that
\begin{equation}
\psi(L)=A L^{\alpha}10^{\beta[log(L)]^{2}}
\end{equation} 

We refer to $\alpha$ as the power-law index and $\beta$ as the curvature index. A large 
$\alpha$ value ($\alpha \ge 0$) indicates that the burst has a large amount of 
high luminosity emission relative to low luminosity emission, while a small value of 
$\alpha$ ($\alpha \ll 0$) indicates that the burst has a significant amount of emission at 
luminosities less than the peak luminosity. 
Simply put, a large $\alpha$ value indicates that a GRB spends little time emitting
below the peak intensity, whereas a small $\alpha$ value indicates that a GRB spends 
considerable time emitting below the peak intensity. In other words, a small $\alpha$ value
indicates {\em persistent} low-intensity emission. 

The introduction of $\beta$ represents a significant fitting improvement over the
use of $\alpha$ alone by \citet{hor97}. The $\beta$ parameter is introduced because some
GRBs have $\alpha$ values that change significantly as the threshold is decreased from
$3\sigma$ to $1\sigma$ above background, whereas $\alpha$ does 
not change for other GRBs . The addition of $\beta$ resolves this fitting problem while simultaneously
indicating that $\beta$ represents a distinguishable ILF property that only affects certain
bursts. A large value of $\beta$ ($\beta \ge 0$) indicates that there 
is no curvature such that the ILF is fitted by a simple power law. A small $\beta$ 
($\beta \ll0$) indicates that the burst is also depleted in low luminosity emission relative to 
that expected from the power-law index $\alpha$. The quantities $\alpha$ and $\beta$ are highly correlated for GRBs; 
large $\alpha$-values indicate large $\beta$-values, and small $\alpha$-values 
indicate small $\beta$-values. In the distribution function, $\beta$ indicates
whether or not the intensity trend found from $\alpha$ continues to lower luminosity 
emission. The correlation between $\alpha$ and $\beta$ demonstrates that bursts with
a large amount of emission just below the peak intensity will also have less low-intensity
emission than expected. More detailed discussions of how pulse shape and the distribution of pulses affects the ILF can be found in \citet{hor97, hak03b, hak04b, hak06}.  Typical uncertainties in measuring $\alpha$ and $\beta$ are $\sigma_{\alpha} \approx 0.7$ and $\sigma_{\beta} \approx 0.8$.

\subsection{Excluding the Short GRB Class}

We have chosen to concentrate our analysis on the Long class of GRBs, as there is 
ample evidence that the Short class represents a different physical mechanism than the Long 
bursts  \citep{kou93, bal04, hjo05, nor06}. 

A detailed analysis to delineate Long and Short GRBs has been previously performed on BATSE data using four different statistical clustering algorithms \citep{hak03}. Each algorithm easily identified the Short burst class using burst attributes of 50-300 keV fluence, T90 duration (the time between recording 5\% and 95\% of the burst emission), and HR321 hardness ratio (100 to 300 keV fluence divided by the 25 to 100 keV fluence). Although the class characteristics and membership differ slightly based on the algorithm, as well as on the choice of attributes, the algorithms were all able to identify a well-defined group of bursts with short T90s, small fluences, and large hardness ratios.  By assuming Gaussian population distributions and by simplifying the classification by removing fluence, the EM algorithm \citep{dem77} allowed the following classification rules to be produced for Short bursts \citep{hak03}:  (T90 $<1.954$) OR ($1.954 \le T90 < 4.672$ AND HR321 $> 3.01$). The following analysis thus excludes GRBs with these characteristics.

\subsection{Overall Dataset}

The results quoted here are limited to the highest-quality ILF measurements; e.g. those 
for which the ILF has at least six degrees of freedom. Our sample contains 1482 GRBs with measured channel 31 lags and high quality ILFs measured in the 50 to 300 keV range above the $1\sigma$ threshold. Of these, 811 have positive lag measurements. This sample is further reduced by limiting it to the 749 bursts in this sample that have hardness ratios and durations needed to classify them as Long or Short bursts. Of these, 668 are classified as Long bursts. Due to the constraints mentioned here, the sample slightly favors brighter GRBs (as measured by 1024 ms peak flux) with longer durations, but still includes many GRBs near BATSE's detection threshold.

\section{Analysis}
\label{sec-anal}

\begin{figure}
\plotone{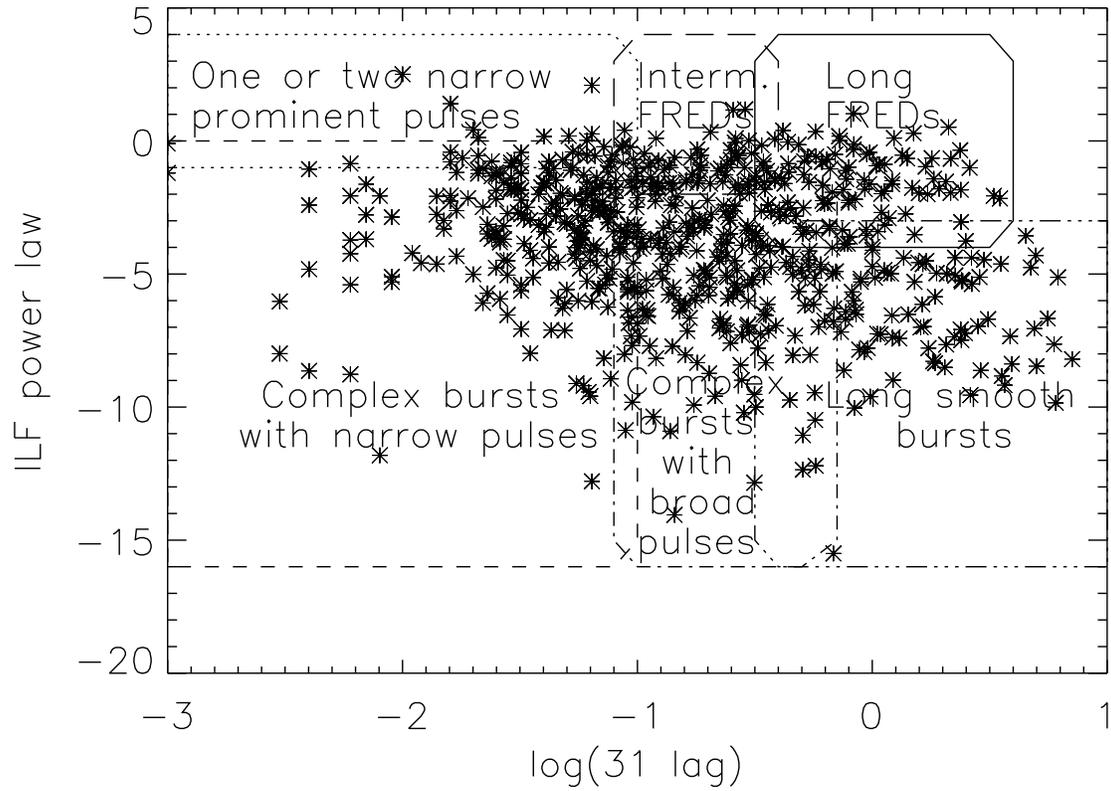}
\caption{ILF power law index vs. 31 lag. Regions are identified in which GRBs are found to have similar time history morphologies.\label{fig1}}
\end{figure}

Figure~\ref{fig1} is a plot of $\alpha$ vs. $\log(\rm lag_{31})$ for the 
sample. It is not surprising that GRBs with different time histories occupy different regions of the plot, since the ILF and the lag describe essentially independent characteristics of the time history. In fact, this figure is a roadmap to GRB time histories, as bursts found in
different regions of the plot generally have similar time histories. To demonstrate this, 
we show six selected time histories from each of six `regions'.

\begin{figure}
\plotone{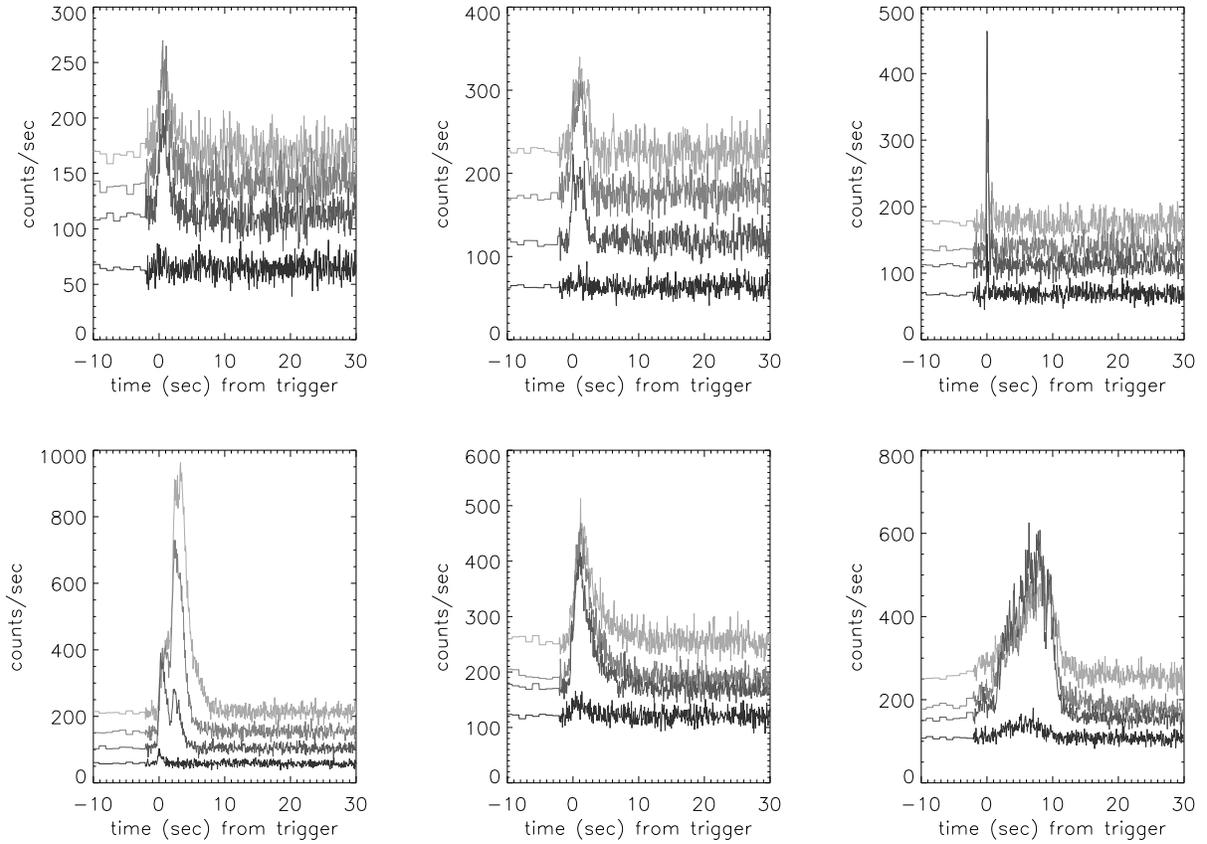}
\caption{Sample time histories of GRBs selected on the basis of large
ILF power-law indices and short lags. These are typically short, peaked, 
bursts dominated by one or two prominent pulses. The Short class of GRBs
also occupies this region, but these bursts have been excluded from our
study. \label{fig2}}
\end{figure}

\begin{figure}
\plotone{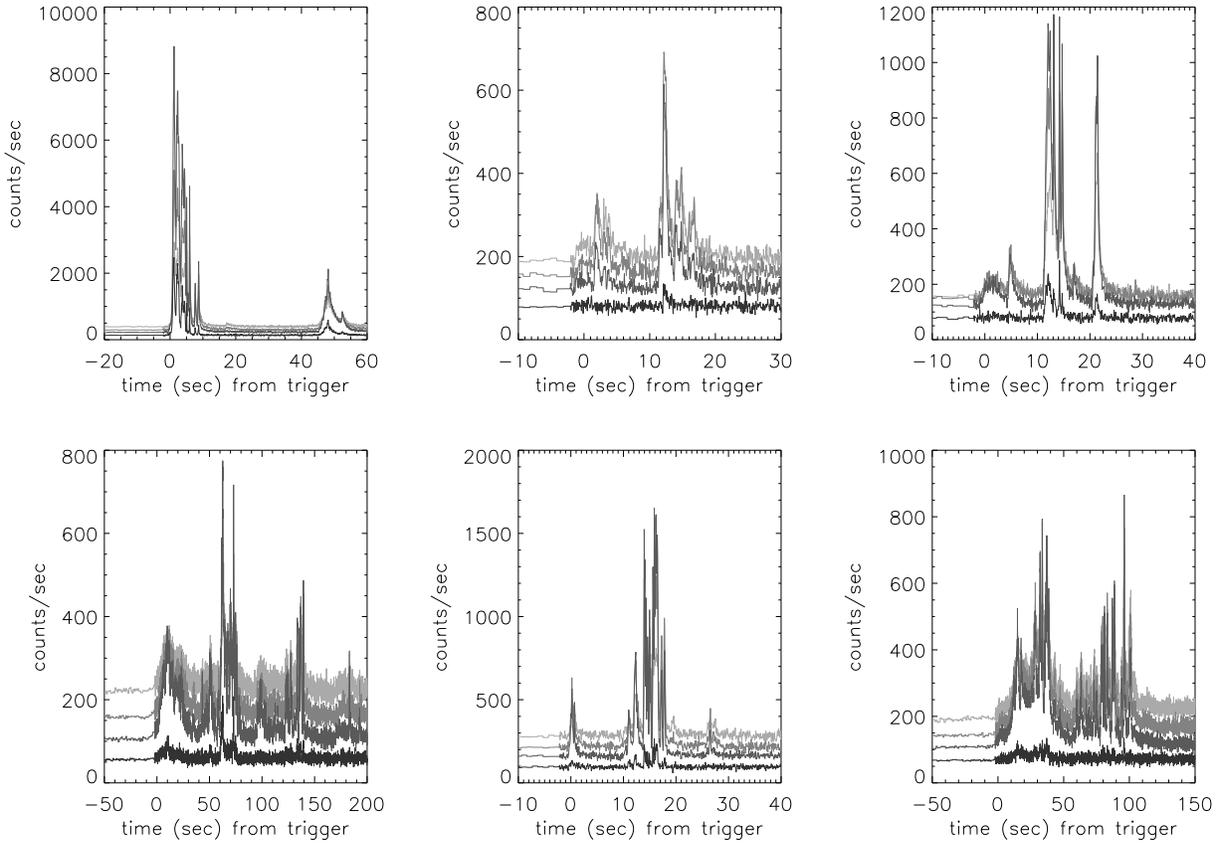}
\caption{Sample time histories of GRBs selected on the basis of small
ILF power-law indices and short lags. These bursts are typically
characterized by complex time histories containing many narrow pulses.\label{fig3}}
\end{figure}

\begin{figure}
\plotone{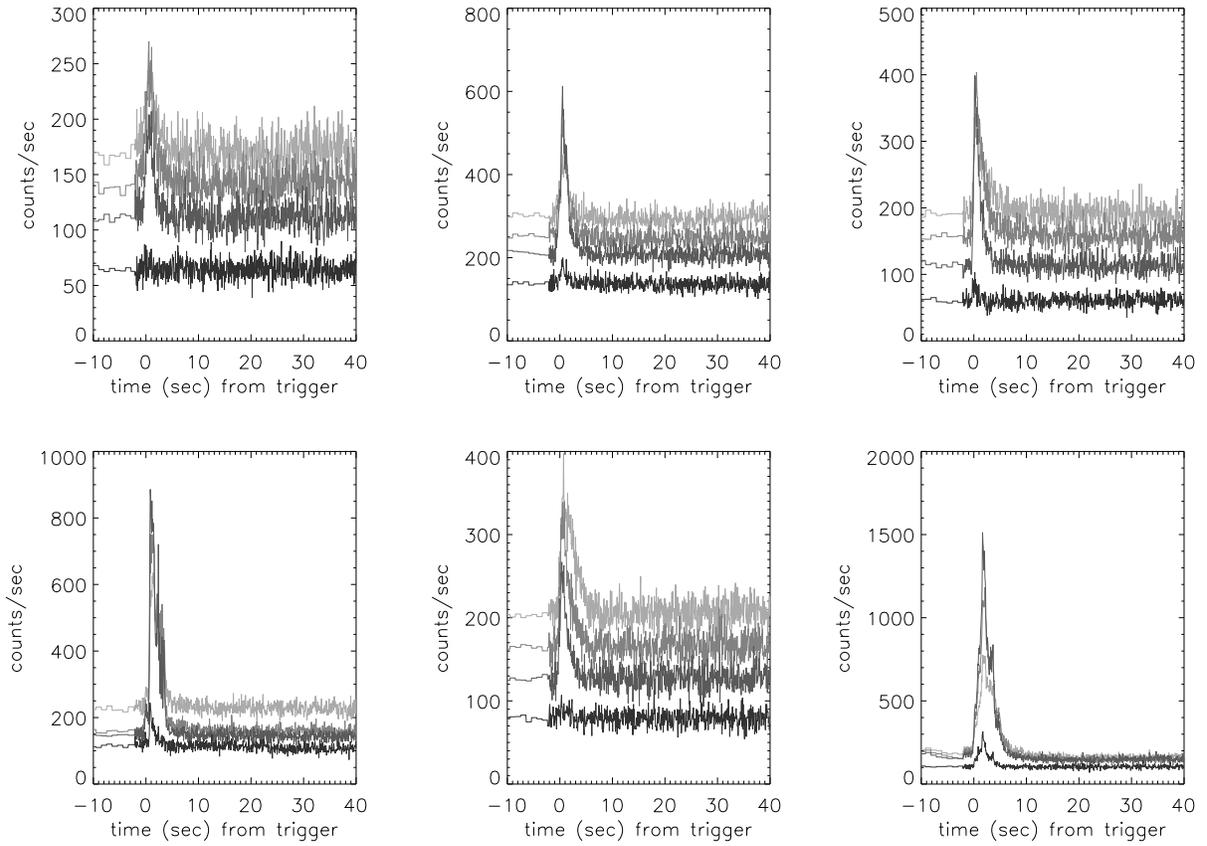}
\caption{Sample time histories of GRBs selected on the basis of large
ILF power-law indices and moderate lags. These bursts are typically 
characterized by FREDs of intermediate duration.\label{fig4}}
\end{figure}

\begin{figure}
\plotone{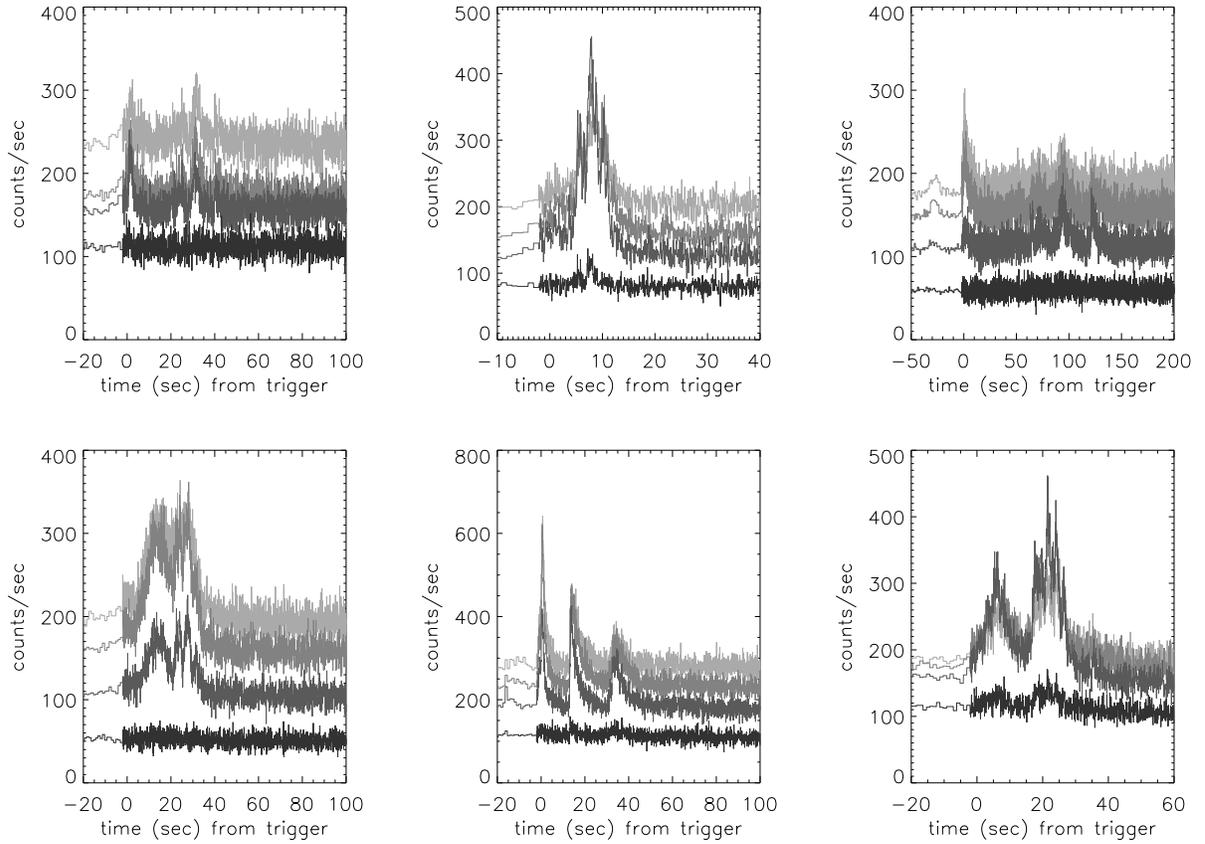}
\caption{Sample time histories of GRBs selected on the basis of small
ILF power-law indices and moderate lags. These bursts are typically characterized
by complex time histories containing many pulses of intermediate duration.\label{fig5}}
\end{figure}

\begin{figure}
\plotone{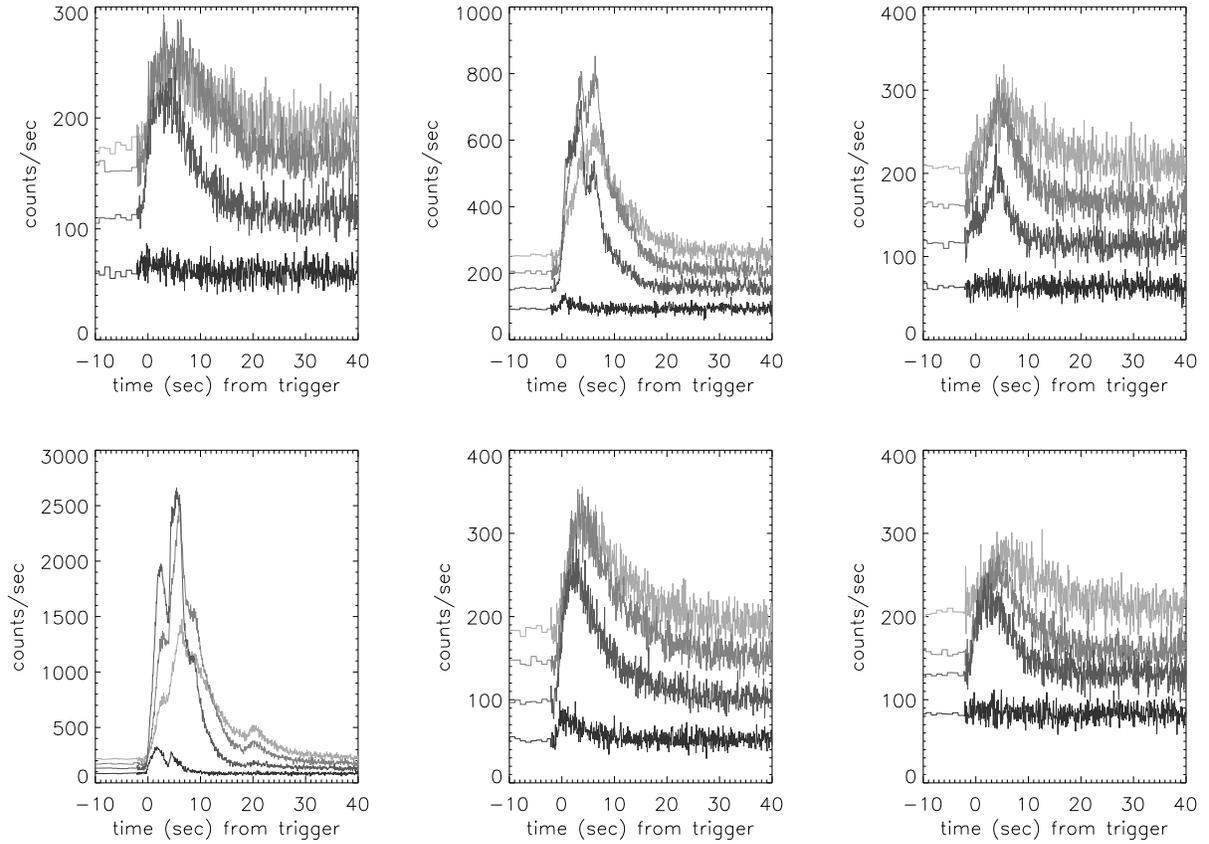}
\caption{Sample time histories of GRBs selected on the basis of large
ILF power-law indices and long lags. These are typically long, peaked, 
smooth FRED bursts.\label{fig6}}
\end{figure}

\begin{figure}
\plotone{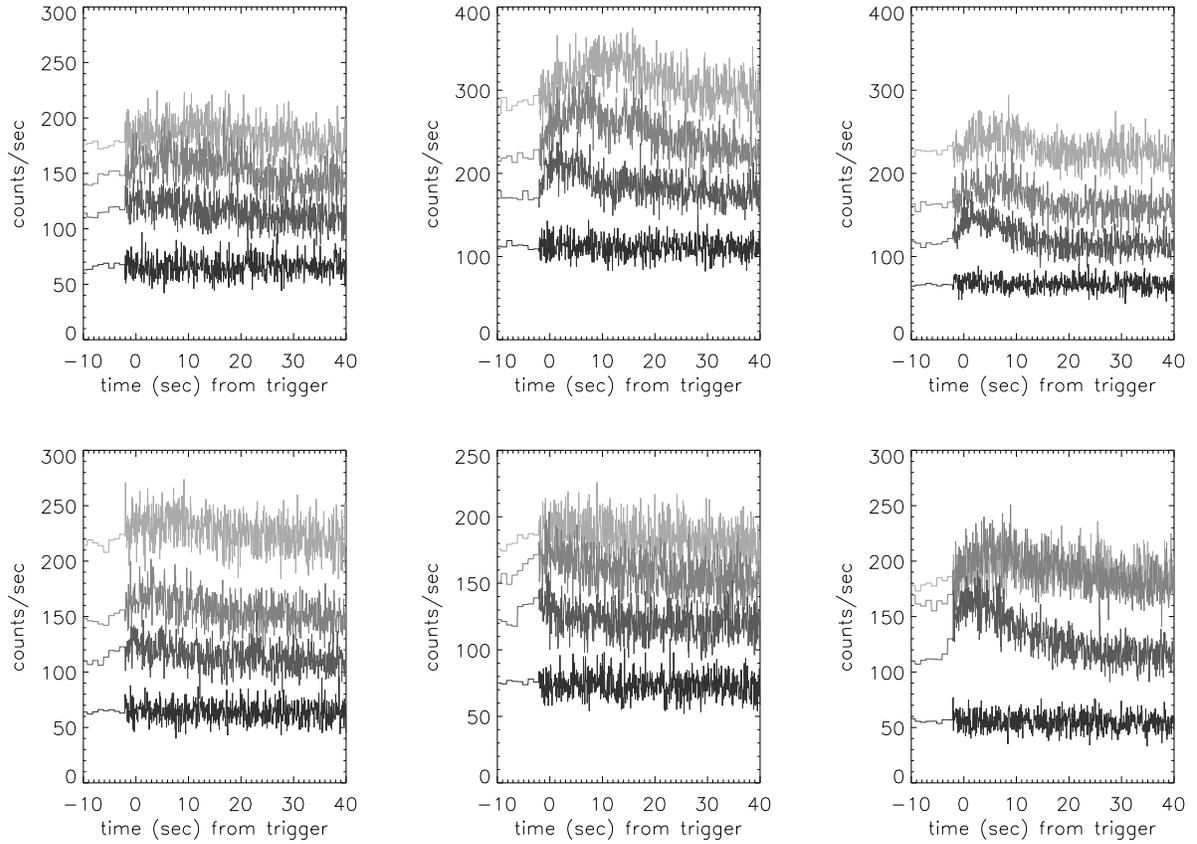}
\caption{Sample time histories of GRBs selected on the basis of small
ILF power-law indices and long lags. These bursts generally exhibit a single 
broad, unpeaked smooth pulse.\label{fig7}}
\end{figure}

Figures~\ref{fig2} through \ref{fig7} show representative GRB morphologies of Long
bursts found in each of the corresponding regions. 
Figure~\ref{fig2} shows representative bursts having short lags and large ILF indices ($\alpha \approx 0$). Bursts in this region typically have short durations and are dominated by one or two pulses. This region is underpopulated compared to the other regions, although the population is increased dramatically when Short bursts (typically short duration with few pulses) are included in the sample.
Figure~\ref{fig3} demonstrates sample
time histories from the short lag, small ILF index region. The small ILF 
power-law index values ($\alpha \ll 0$) indicate that these GRBs spend considerable time 
emitting at luminosities below the peak luminosity. Bursts with small ILF power-law
indices are also found to have very negative curvature indices. Both of these characteristics can arise when multiple pulses are present: an increased number of fainter pulses can increase the faint component of the distribution function. These bursts also tend to have a significant amount of hard emission. 
Figure~\ref{fig4} shows bursts that are similar to those in Figure~\ref{fig2} in that
they have large ILF indices ($\alpha \approx 0$), but which have intermediate lags. These bursts are typically 
FREDs of intermediate pulse width.
Figure~\ref{fig5} demonstrates the time histories of bursts with short/intermediate lags and small ILF 
indices ($\alpha \ll 0$). Typical bursts with these characteristics are complex with several broad pulses.  
Figure~\ref{fig6} identifies GRBs having long lags and large ILF indices ($\alpha \approx 0$). 
Bursts in this region are typically long duration FREDs. 
Figure~\ref{fig7} identifies GRBs having long lags along with small ILF indices ($\alpha \ll 0$). 
Bursts in this region are cuspless, long, smooth, single-pulsed bursts (Long smooth bursts) which have small ILF indices because their pulses do not have identifiable peaks. This causes the bulk of their emission to be, unlike FREDs, close in intensity to the peak luminosity. A well-known example of this morphological type is GRB 980425, the underluminous burst associated with SN 1998bw. The secondary pulses of quiescent GRBs 960530 and 980125 \citep{hak04} have pulse shapes consistent with this GRB morphology, indicating that the diagram can also serve as a {\em pulse} morphology indicator.

Based on Figure~\ref{fig1}, and verifying some of our observations with a small database containing estimated numbers of GRB pulses \citep{ful03}, we summarize the relationship between various general GRB time history characteristics as obtained from lags and ILFs as follows:
\begin{itemize}
\item{GRBs with longer lags have fewer, broader pulses than those with shorter lags. For example,
complex bursts with broad pulses typically have half as many measured pulses as complex
bursts with narrow pulses.}
\item{Small $\alpha$ values (corresponding to a decrease in the fraction of high-luminosity to low-luminosity emission) for short- and intermediate-lag bursts indicate a larger number of pulses.}
\item{Long-lag GRBs have two morphological types: single-pulsed FREDs and Long, smooth bursts (these have also been identified by \citet{nor05}). For long-lag GRBs, $\alpha$ primarily measures the change in pulse shape; large $\alpha$ values identify single-pulsed FREDs while small $\alpha$ values identify Long, smooth bursts. There are a few long-lag GRBs with overlapping properties indicating that the two morphologies are not distinct. The sensitivity of the ILF to the number of pulses and to pulse shape has been noted previously  \citep{hor97}. We also note here that the effect of pulse shape on the ILF becomes increasingly noticeable at longer lags because long-lag single-pulsed bursts can be found with smaller $\alpha$ values than short-lag single-pulsed GRBs (see Figure \ref{fig1}). There do not appear to be short-lag Long, smooth bursts.}
\end{itemize}

\section{Discussion}

\subsection{The Lag and the ILF as Intrinsic GRB Attributes}

We now show that Figure~\ref{fig1} primarily identifies {\em intrinsic} rather than {\em extrinsic} GRB parameters. This statement means that intrinsic characteristics such as bulk Lorentz factor and jet beaming angle dominate over cosmological effects. Here the term ``intrinsic'' means that the observed quantity is not significantly affected by cosmology or by other observational effects.

\subsubsection{The Lag as a Primarily Intrinsic Attribute}

Lag is an intrinsic attribute that anti-correlates with burst peak luminosity (long-lag bursts are less luminous than short-lag bursts in the comoving frame; \citet{nor00}). The lag vs. luminosity relation indicates that short-lag bursts are luminous and are typically found at large redshift. All other effects being equal, cosmological time dilation would tend to cause the most distant, most luminous GRBs to have the broadest pulses and the longest lags.
Cosmological time dilation should also make intermediate- and long-lag bursts into stretched-out versions of short-lag bursts with similar ILF values.  Instead, long-lag bursts typically have fewer pulses than short-lag bursts, indicating that the two burst types are inherently different. Finally, the overlap in lag between GRB morphological types is less than one might expect. By this, we mean that the vast majority of bursts with a specific morphological type (such as narrow-pulsed bursts with complex time profiles) have lags that delineate them from bursts of a different morphology (such as intermediate-pulse width bursts with complex time profiles). If cosmological time dilation were as important as the burst's intrinsic characteristics, then one would expect such an overlap in lag between the two morphological types that the delineation would be completely smeared out. This is not observed. 

Weaker evidence that lag is intrinsic and is correlated with morphological types comes from a study of GRB Auto Correlation Functions (ACFs; \citet{bor04}).  A bimodal distribution of ACF half-maximum widths has been identified; a weak correlation at the 20\% significance level) can be found between these widths and $\rm lag_{31}$ in the GRB rest frame. Broad ACFs typically indicate broad-pulse burst components (e.g. \citet{ban97, bel00, rrm00}). The separation between narrow and broad ACF distributions possibly corresponds to the lag separation between complex bursts with broad pulses and complex bursts with narrow pulses in Figure~\ref{fig1}; this could represent an intrinsic delineation between narrow- and broad-pulse events. 

\subsubsection{The ILF as a Primarily Intrinsic Attribute}

GRBs with differing $\alpha$ values but similar lags have intrinsically different pulse structures. The number of pulses and the pulse shape are intrinsic characteristics that describe how we observe a burst's energy release. For short- and intermediate-lag GRBs, the ILF power-law index $\alpha$ is an indicator of the {\em number of burst pulses}. In this manner, it {\em anti-correlates} with burst variability, which is a luminosity indicator \citep{rei01}. For long lag, single-pulsed GRBs, $\alpha$ is a {\em pulse shape indicator} and {\em correlates} with burst variability. Thus, there is a suggestion that the ILF, through variability, might be a luminosity indicator. However, it should be pointed out that variability most strongly correlates with lag, since GRBs with short lags are typically more complex than those with longer lags. Just as for lag, there is relatively little overlap between morphological types having different ILF values. 

We can study $\alpha$ as a potential luminosity indicator using the few BATSE GRBs with measured isotropic luminosities. Table~\ref{tbl-1} summarizes prompt and afterglow luminosities of these bursts. The most luminous BATSE bursts have short lags and large $\alpha$ values, while the least luminous bursts have long lags and small $\alpha$ values. The sample is unfortunately too small to determine if a relationship exists between $\alpha$ and isotropic luminosity for bursts having similar lags.

Single-pulsed GRBs differ from complex GRBs in that they either have large ILF $\alpha$-values or long lags. Single-pulsed bursts are interesting for two reasons: (1) there is statistical evidence that GRBs associated with supernovae tend to have simple single- or double-pulsed time histories \citep{bos06}; these associations include the direct associations of GRB 980425 (a Long smooth burst) with SN1998bw, GRB 031203 (a long FRED; \citet{saz04}) with SN2003iw \citep{mal04}, and GRB 030329 (a double-pulsed FRED; \citet{van04}) with SN2003dh \citep{hjo03}, and (2) a distinct class of GRBs with low-luminosity afterglows has recently been identified; these bursts are predominantly single-pulsed \citep{lia06, nar06}. 

Thus, single-pulsed bursts typically exhibit a consistent pulse shape (the FRED) that remains ubiquitous over a wide range of lags, or they exhibit FRED or Long smooth pulse shapes if they have very long lags. The types of supernovae associated with GRBs and the class of low-luminosity afterglows both appear to be dependent upon these constraints.

\subsubsection{The Impact of Cosmological Redshift on Spectral-Dependent Lag and ILF Measurements}

One might be concerned that cosmology, through the influence of redshift on burst spectra, could play a role in altering either lags or ILF measurements significantly from their rest frame values. However, this does not seem to be the case. We note that lags deviate only slightly across energy channels in the 25 keV to 300 keV range, as can be seen from the statistical relationship that

\begin{equation}
{\rm lag_{21}} + {\rm lag_{32}} = -0.0032 + 0.98 ({\rm lag_{31}}) - 0.026 ({\rm lag_{31}})^2.
\end{equation}

In the absence of nonlinear spectral effects, one would expect the linear relationship $\rm lag_{21} + \rm lag_{32} = \rm lag_{31}$; the observed relationship is indeed very close to this. Furthermore, if the nonlinearity were due to cosmological effects, then most of the nonlinearity should occur for the short-lag, high redshift bursts, rather than for the long-lag, low redshift bursts for which it is observed. This is because the long-lag GRBs are at low redshift, which places them nearby where nonlinear effects should be less pronounced.

The impact of cosmological redshift on the ILF is more complicated. The primary energy-dependent effect on the ILF results because pulses are broader at low than at high energies \citep{nor96}. We have demonstrated in Section~\ref{sec-anal} and Figure~\ref{fig1} that broader single pulses produce smaller $\alpha$ values than narrow pulses; this means that low-energy $\alpha$ values should typically be smaller than high energy $\alpha$ values. An analysis of the $\alpha$ values obtained for bursts in various BATSE energy channels shows that this is indeed the case. Cosmological redshift effects must be compared to pulse broadening effects. Cosmological redshift causes photons emitted at particular rest frame energies to be observed at lower energies in the observer's reference frame. The pulses of high redshift bursts are thus expected to be narrower than the pulses of low redshift bursts observed in the same energy channel.  If we assume that luminous GRBs with small lags are observed at a greater redshift than low-luminosity GRBs with large lags, then we might have two samples for which we could study the cosmological energy dependence in greater detail. However, the relationships between $\alpha$ values observed in different energy channels are nonlinear, and therefore difficult to interpret. There could be many reasons for this. For example, the change in pulse shape for long lag bursts could be responsible. Without a better model describing how and why this change occurs, we are not in a position to interpret the nonlinearity of the observations. Thus, we cannot comment on the impact of cosmological redshift on the ILF other than to say that it appears to be of secondary importance to the effect of energy-dependent pulse broadening. 

\subsection{Tying the Observations into GRB Models}
\label{sec-model}

GRB time history morphologies appear indicative of both isotropic gamma-ray luminosity and optical/infrared afterglow luminosity. A relatively simple model can explain these characteristics. 

Lag is an indicator of jet opening angle \citep{fra01, pan02, nor02} and of Lorentz factor \citep{she05}; a short lag identifies a tightly-beamed jet with a large Lorentz factor while a long lag is more indicative of a broadly-beamed jet with a smaller Lorentz factor.  The lag also appears to depend on the cooling time in the comoving frame and the low-energy spectral index, but not on the fireball radius \citep{she05}.  It follows that a large Lorentz factor should deposit a larger fraction of the bulk energy into the afterglow than a small Lorentz factor, resulting in an anti-correlation between lag and isotropic burst luminosity. An anti-correlation of this type can be seen in the Table~\ref{tbl-1} data.

The ILF indicates the {\em number of pulses} for short-lag GRBs and the {\em pulse shape} for long-lag GRBs. The single-pulsed Long, smooth bursts have the largest opening angles, lowest gamma-ray and afterglow luminosities, and likely the smallest Lorentz factors of any GRB type. They are also lacking in detected emission above 300 keV.  It is thus not surprising that these bursts appear to deposit the least energy of any GRB type into the afterglow \citep{lia06, nar06}. However, Long GRBs with narrow single pulses (and shorter lags) also produce low luminosity afterglows (e.g HETE GRBs 021211 and 040924; \citet{don06}).

\begin{figure}
\plotone{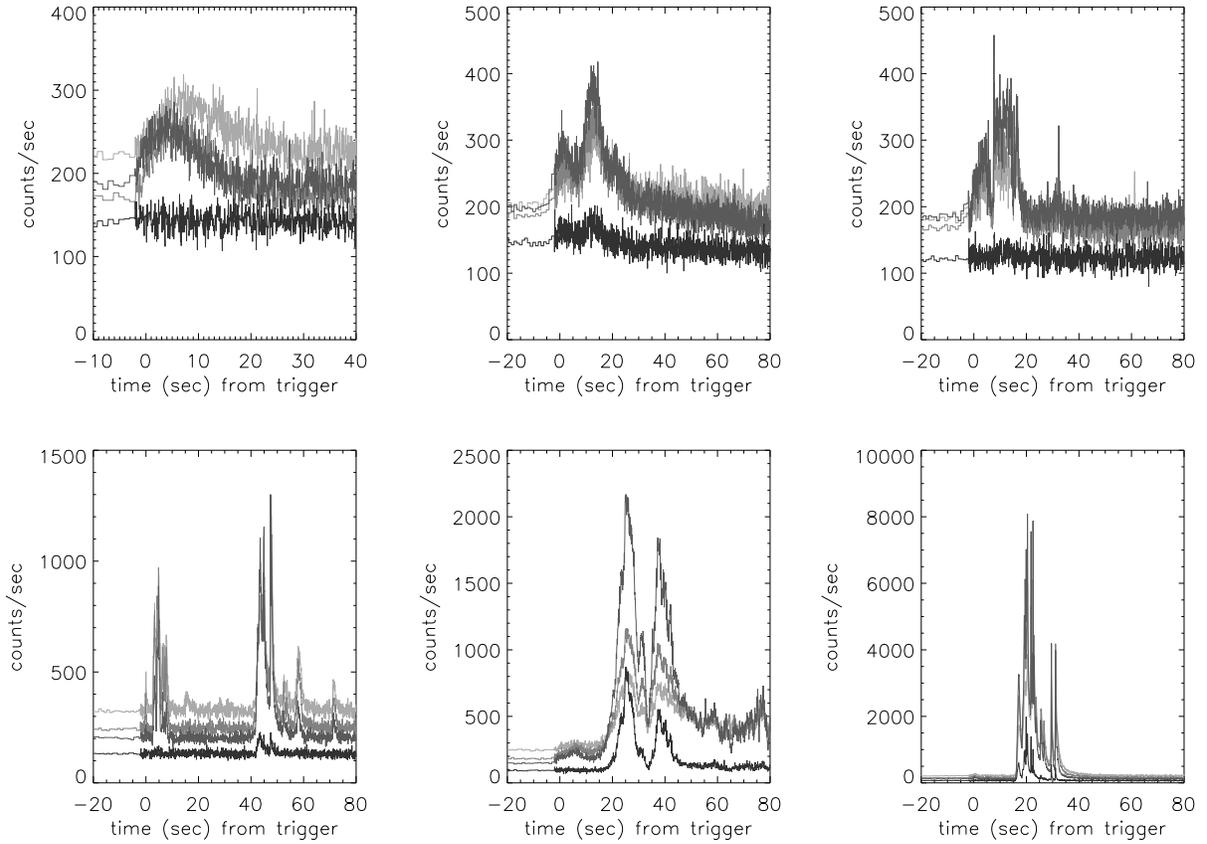}
\caption{Time histories of BATSE GRBs in Table~\ref{tbl-1}. From top left to lower right: low afterglow luminosity GRB 980425, and high luminosity afterglow GRBs 980703, 971214, 990510, 990123, and 991216.\label{fig8}}
\end{figure}

\begin{figure}
\plottwo{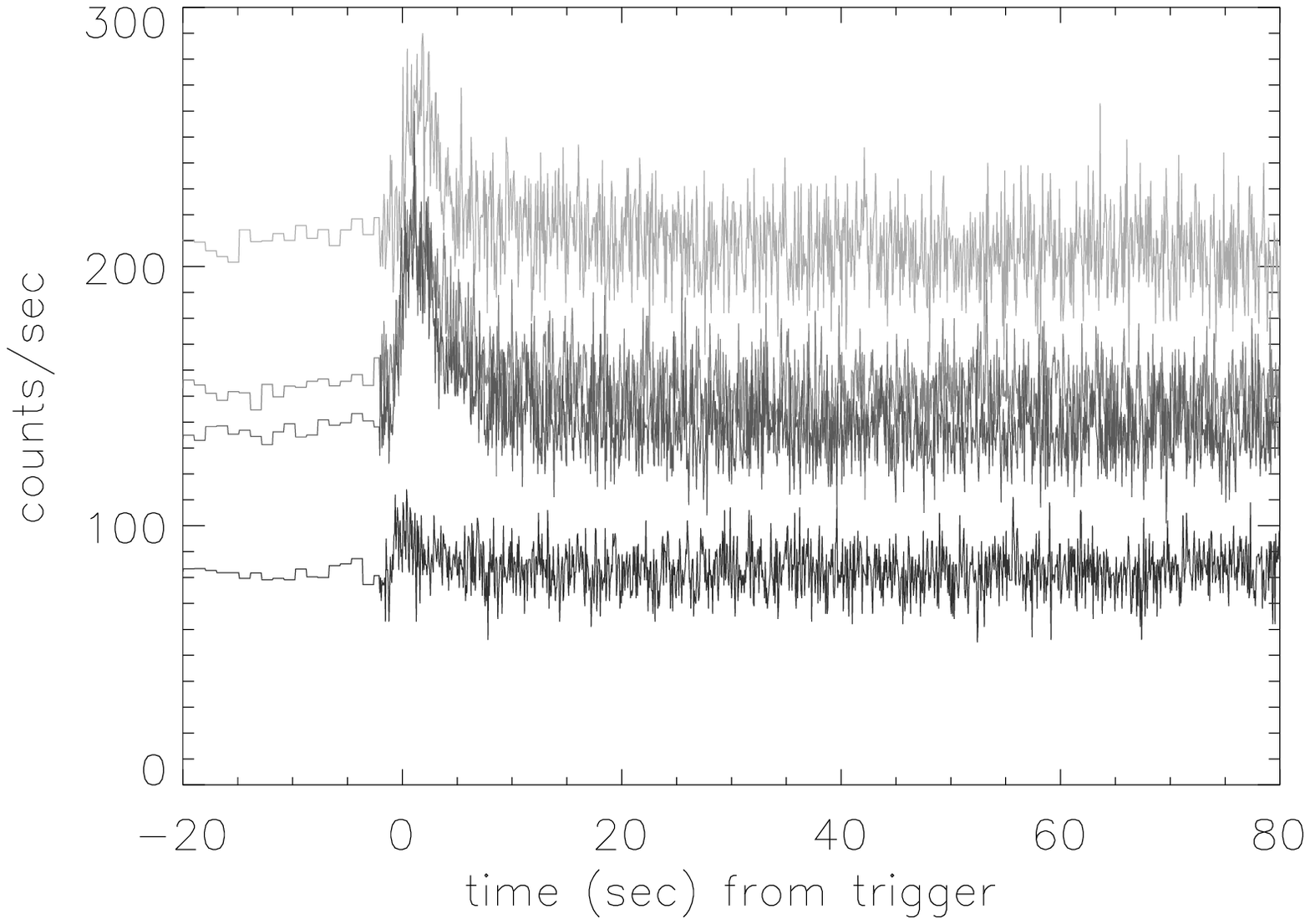}{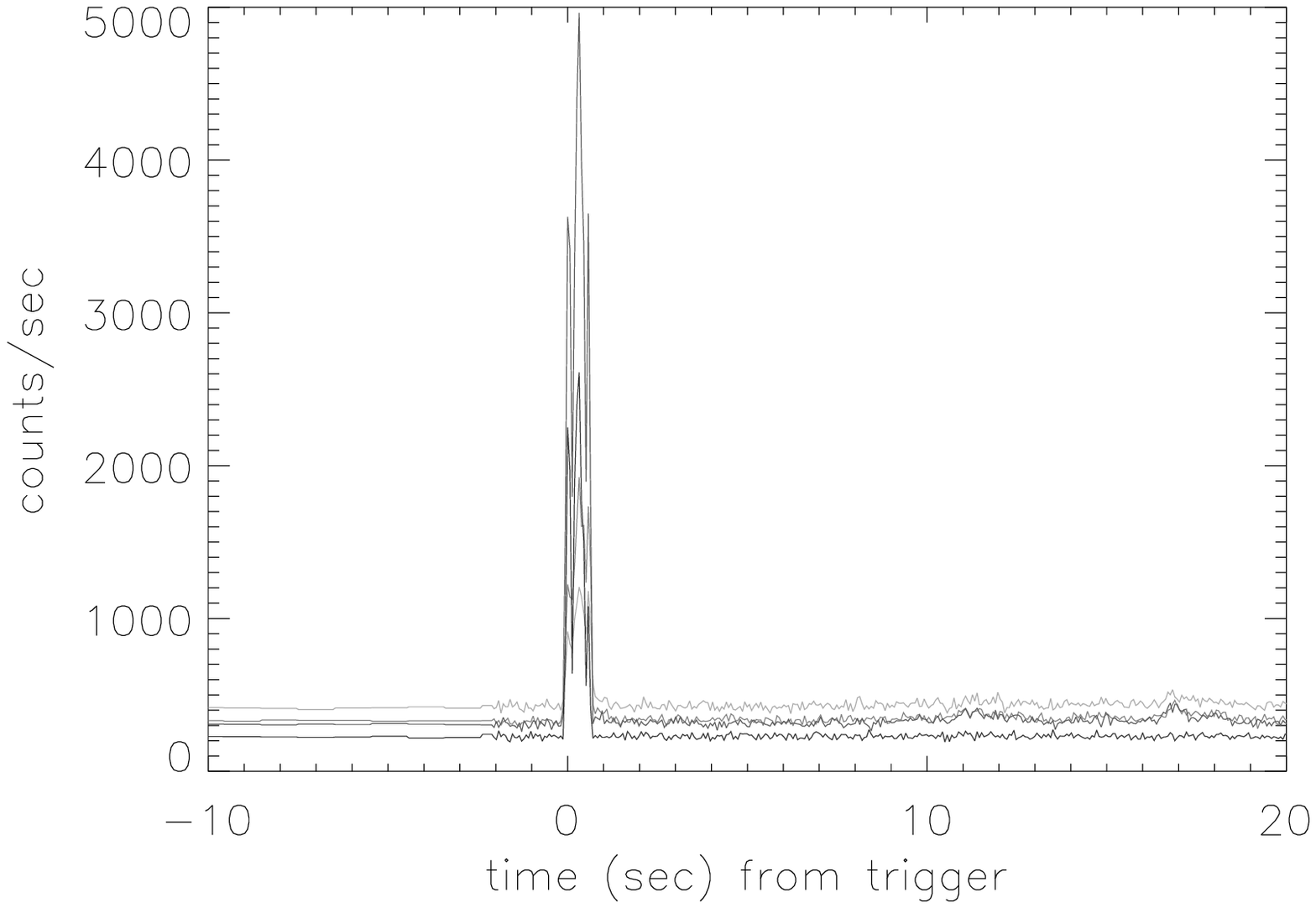}
\caption{Time histories of BATSE GRBs 970508 (left) and 990712 (right). In Section~\ref{sec-model} we suggest that 1) additional pulses have occurred for GRB 970508 during the afterglow phase, and 2) GRB 990712 be reclassified as a Short burst. With these caveats, the time histories, ILFs, and lags of these GRBs are all consistent with the single vs. multiple pulse model described in the text.\label{fig9}}
\end{figure}

The two optical/infrared afterglow classes found by \citet{lia06} and \citet{nar06} appear to result from the discrete distinction between GRBs with a single prominent pulse and those with more than one prominent pulse. From an energetics perspective, this might have to do with the quantized amount of energy available in a single gamma-ray pulse. The majority of BATSE GRBs with high luminosity afterglows listed in Table~\ref{tbl-1} are multi-pulsed (see Figure~\ref{fig8}), while low luminosity GRB 980425 is a single-pulsed Long smooth burst. However, two exceptions warrant reexamination: GRB 970508 (which \citet{lia06} list as having a single pulse and a high afterglow luminosity) and GRB 990712 (which \citet{lia06} list as having many pulses and a low afterglow luminosity). The time histories for these two GRBs are shown in Figure~\ref{fig9}.

GRB 970508 is single-pulsed high optical/infrared luminosity afterglow burst that starts out with a flat afterglow light curve and a low luminosity, then evolves into a luminous afterglow with significant fluctuations \citep{ped98}. These afterglow characteristics have been attributed to additional energy injections during the afterglow phase \citep{dai01, bjo04, fox03, zha05}, which \citet{lia06} use to support its placement on the high-luminosity branch. The additional energy injections have either occurred at late times or (more likely) at energies below the instrumental threshold so that they were not observed by BATSE. This means that the pulses presumed to be associated with these injections have not been included in either the morphological description or the ILF measurement. If the pulses had been included, then GRB 970508 would have been treated as a multi-pulsed GRB with a steeper ILF, and would been consistent with other bursts found in the high-luminosity afterglow branch.

Although GRB 990712 appears to be a single-pulsed burst, closer examination (e.g. Figure~\ref{fig9}) indicates that two long, faint pulses follow the obvious bright initial spike, and that that this spike contains three very short evolving pulses. The overall lag of this burst is short (0.045s +/- 0.014s), and the ILF is not measurable using the procedures that we have implemented. The ILF cannot be measured because the fluence is primarily contained in the main pulse (with too few temporal bins for an accurate measurement), while the burst duration is defined by the extended emission (with too few intensity bins for an accurate measurement). Only eight GRBs in the BATSE Catalog are known to have a narrow initial spike followed by faint extended emission; these are thought to be Short bursts \citep{nor06}. This interpretation is consistent with the morphologies identified by our roadmap; GRB 990712 does not have the appearance of the short-lag FREDs typically found in the same lag vs. ILF region of Figure~\ref{fig1}. GRB 990712 was included in our Long burst sample because its T90 (used as a classification attribute) was consistent with that of a Long burst. However, based on this reexamination, we concur with the interpretation of \citet{nor06} and subsequently exclude GRB 990712 from our sample.

Thus, there is evidence to suggest that single-pulsed GRBs with large ILF values might have low luminosity afterglows, whereas their multi-pulsed counterparts with similar lags might have small ILF values and high luminosity afterglows. We hypothesize that this indicates the difference in discrete energies available to be transmitted from the external shock into the afterglow via one pulse or many pulses.

Therefore, 1) single- and multi-pulsed bursts with similar lags have similar jet beaming angles, yet single-pulsed FREDs typically deposit less energy than the multi-pulsed bursts into the afterglow \citep{lia06, nar06}, 2) there is an association between Type Ibc supernovae and single-peaked Long GRBs having soft spectra and smooth time histories \citep{bos06}, and 3) long-lag GRBs are single-pulsed events with the lowest luminosities and little high-energy emission. When taken together, the observations suggest that single-pulsed GRBs release the bulk of their energy in a single blast, similar to a supernova. This similarity may be more noticeable for long-lag bursts whose energy release is more isotropic.

\begin{figure}
\plotone{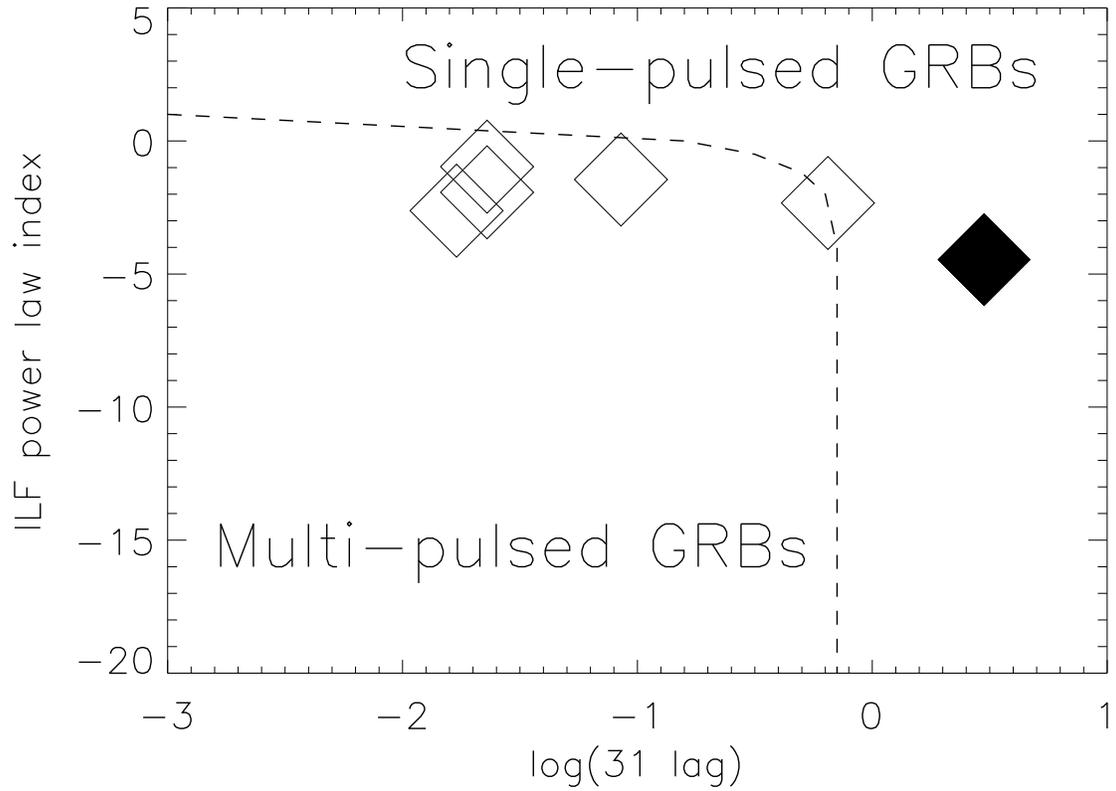}
\caption{BATSE GRBs with known luminosities and optical/infrared afterglow luminosities. A rough dividing line between single- and multi-pulsed GRBs observed by BATSE is shown (dotted line). 
Low-luminosity afterglow GRB 980425 is indicated by a filled diamond, while multi-pulsed GRBs are identified by open diamonds.\label{fig10}}
\end{figure}

In Figure~\ref{fig10}, we have plotted the ILF $\alpha$ and $\log(\rm lag_{31})$ values of the remaining BATSE bursts from Table~\ref{tbl-1}. We have identified the rough region from Figure~\ref{fig1} occupied single-pulsed bursts. The lone single-pulsed GRB 980425 has been denoted by a filled diamond, to indicate that it belongs to the class of bursts having low luminosity afterglows, while the other bursts having high-luminosity afterglows have been identified by open diamonds.

If the strong connection between number of pulses, afterglow luminosity, and supernova connection turns out to be valid, then this would suggest that the number of pulses found in a GRB is an inherent property (solely reflective of the central engine activity plus Lorentz factor) rather than an emergent one (depending on a combination of environmental factors, including viewing angle, plus the Lorentz factor). 
Although environmental factors such as viewing angle and/or filling factor might cause the observed number of pulses to change without otherwise changing the relativistic flow, these factors should not correspondingly change the bulk energy deposited into the external shock.

In summary, Figure~\ref{fig1} constrains theoretical GRB models because there is an apparent association between beamed prompt emission and delayed, low-energy, more isotropic emission: simple GRB time histories are correlated with low-luminosity afterglows and/or supernova light curves, while complex GRB time histories generate luminous afterglows and do not appear to be associated with supernovae. In other words, a complex relativistic shock signature produces one type of delayed emission while a simple shock signature produces another type of delayed emission. Due to geometry, multi-pulsed bursts must be seen at small viewing angles (e.g. \citet{sar97}), {\em and yet multi-pulsed GRBs are not observed to have low-luminosity afterglows.} This argues that the relationship between single-pulsed GRBs and low-luminosity afterglows is likely intrinsic. Thus, GRB time history morphologies may represent a more complete record of GRB physics than has previously been suspected.

\subsection{Other Applications of the GRB Morphology Roadmap}

Figure~\ref{fig1} can be used to study other properties of morphologically similar GRBs. Two simple applications are listed here: 

\begin{itemize}
\item{Properties of morphologically similar GRBs can be studied upon selecting GRB types
on the basis of the lag and ILF values. For example, we verify that 
long-lag FREDs have longer T90 durations than short-lag FREDs. A sample of 27
FREDs found on the basis of $\alpha$ and $\rm lag_{31}$ alone show a direct correlation 
between $\rm lag_{31}$ and T90; the probability is only $p=1.8 \times 10^{-4}$ that the 
correlation is random. However, a low-dispersion linear fit to this correlation
is not good, as a few of these GRBs have faint secondary pulses that
dramatically increase their T90s without significantly influencing their ILF power-law
indices. We thus conclude that $\rm lag_{31}$ is a more consistent temporal indicator 
of FRED properties than is T90; this result is in agreement with the conclusions of \citet{don06} from a study of HETE GRBs.}
\item{The similarity between GRB morphologies provides 
additional information that can be used to support or reject GRB classification suspected for other reasons. For example, the Short GRB class is defined primarily on the basis of duration and spectral hardness, and the similar ILFs, lags, and time histories of these bursts provide support for this classification. However, narrow-pulsed short events with extended emission such as GRB 990712 appear to belong to the Short GRB class, even though they have long T90s. The morphologies of GRBs belonging to the statistically significant Intermediate class (e.g. \citet{muk98, hor98}) can also be examined. Although the bursts in this class are generally
soft and moderately short, they appear to be composed of a variety of disparate time history
morphologies: this `class' is split between short FREDs and complex bursts with broad pulses.
This argues against the intermediate class being composed of a homogeneous morphological type.}
\end{itemize}

\section{Conclusions}

The Internal Luminosity Function and spectral lags can be used to sort GRBs by their time history morphologies. This produces a roadmap for delineating GRBs morphologically. We are applying data mining tools to the problem to more formally verify whether or not any of these morphological `types' represent true statistical classes.

The morphologies primarily identify intrinsic rather than extrinsic characteristics; this is because internal relativistic effects due to bulk Lorentz factor are much larger than relativistic effects due to cosmology, and because the number of pulses and the shape of pulses indicate different efficiencies and forms of GRB energy release. GRBs with long lags have lower energies and more isotropic emission than GRBs with short lags \citep{fra01,pan02,nor02,she05}. Single-pulsed bursts appear to be more closely associated with supernovae than multi-pulsed bursts. They also appear to convert less jet energy into afterglow luminosity than multi-pulsed bursts.

This suggests that a tradeoff exists between the energy released as radiation (1) during the prompt emission, (2) during the afterglow, and/or (3) through a supernova, if conditions permit one to occur. Each of these represents a different way in which energy from the bulk flow is released.

The approach described here for measuring lags and the ILF can be extended to Swift and other GRB experiments, even though these differ from BATSE in energy, flux, and temporal characteristics. It is important to be aware that low energy injections can be made to the afterglow after the time of the prompt emission (e.g. \citet{bar05,bur05,gra06}), thus changing the ILF. We
find that GRB lags are consistent across a wide range of prompt emission energies; 
(e.g. $\rm lag_{31} \approx \rm lag_{21} + \rm lag_{32}$), so that lags calculated for other
experiments should be scalable to BATSE results. Temporal bin size and instrumental
sensitivity can slightly alter measured ILF values, but we show that these effects can be 
understood in a systematic way \citet{hak06}. To aid future analysis
and discussion, a database of ILF and lag parameters will be placed online alongside a 
large database of GRB attributes and a suite of data mining tools for the purpose of GRB
classification and pattern identification (e.g. \citet{gib04}). This process is already well 
underway \citep{hag05} and may be found at URLs {\tt http://grb.mnsu.edu/grbts/} and {\tt http://grb2.cofc.edu/grbts/}.

\acknowledgments

We are very grateful to referee Jay Norris, whose comments and suggestions improved the manuscript immeasureably. We are indebted to Chris Nolan and Kevin Young for calculating ILFs, to Sarah Sonnett and Christopher Peters for calculating lags, and to Stephen Fuller for calculating ILFs and providing a database containing numbers of GRB pulses. Additionally, we thank Chris Fragile for helpful discussions. We also acknowledge support under NSF grant AST00-98499 and support from South Carolina Space Grant.





\clearpage



\begin{table}
\begin{center}
\caption{Some properties of BATSE GRBs having known redshifts. The properties are compared to two afterglow luminosity classes as identified by \citet{lia06} and \citet{nar06}. In general, single-pulsed bursts appear to have small luminosities as measured from both their prompt emission and their afterglows. The afterglow luminosity classes of GRBs 970508 and 990712 (denoted by asterisks) are discussed in the text. \label{tbl-1}}
\begin{tabular}{lccccccc}
\tableline\tableline
{\bf Burst} & {\bf BATSE} & {\bf $\log L_{\rm Iso}$}\tablenotemark{a} & {\bf $\log L_{R}$}\tablenotemark{b}& {\bf ${\rm lag_{31}}$} & {\bf ILF $\alpha$} & {\bf \# Pulses} & {\bf Afterglow Class}\\
\tableline\tableline
980425  &   6707  &   $48.20$  &   $< 43.9$  &  $2.990 \pm 0.151$  &  $-4.46 \pm 0.78$  &    1      & low\\
990712  &   7647  &   $51.22$  &   $45.02$   &  $0.045 \pm 0.014$  &      ---       &  multi  & low*\\
970508  &   6225  &   $51.85$  &   $45.69$   &  $0.360 \pm 0.083$  &  $-1.71 \pm 0.80$  &    1      & high*\\
980703  &   6891  &   $52.84$  &   $46.76$   &  $0.647 \pm 0.309$  &  $-2.33 \pm 0.35$  &    2      & high\\
971214  &   6533  &   $53.32$  &   $46.30$   &  $0.017 \pm 0.005$  &  $-2.62 \pm 0.49$  &  multi  & high\\
990510  &   7560  &   $53.26$  &   $46.95$   &  $0.038 \pm 0.006$  & $-1.93 \pm 0.24$  &  multi  & high\\
990123  &   7343  &   $54.38$  &   $46.20$   &  $0.085 \pm 0.031$  & $-1.45 \pm 0.32$  &  multi  & high\\
991216  &   7906  &   $53.83$  &   $46.94$   &  $0.023 \pm 0.003$  & $-0.97 \pm 0.19$  &  multi  & high\\

\tableline
\end{tabular}
\tablenotetext{a}{Isotropic gamma-ray luminosities are taken from \citet{nar06}.}
\tablenotetext{b}{Optical/infrared afterglow luminosities at twelve hours are supplied by \citet{lia06b} and are plotted in \citet{lia06}.}

\end{center}
\end{table}

\clearpage

\clearpage

\clearpage

\clearpage

\clearpage

\clearpage

\clearpage

\clearpage


\clearpage

\clearpage

\end{document}